\documentclass[11pt,thmsa,a4paper]{article}
\usepackage[english]{babel}
\usepackage{graphicx,epsf}
\usepackage{lscape,array}
\usepackage[ansinew]{inputenc}
\usepackage{amsthm,amssymb,amsmath,amsfonts}
\usepackage{latexsym}
\usepackage{umlaute}
\renewcommand{\baselinestretch}{1.1}
\def \R {I\!\!R}

\begin{document}

\title{A comparative study of new cross-validated bandwidth selectors for kernel density estimation
}

\author{ \textbf{Enno Mammen\thanks{Corresponding author, Universit{\"a}t Mannheim, Abteilung Volkswirtschaftslehre,
L7, 3-5, 68131-Mannheim, Tel.\ ++49 621 181 1927, FAX ++49 621 181 1931, e mail: emammen@rumms.uni-mannheim.de}},
\textbf{Mar{\'\i}a Dolores Mart{\'\i}nez Miranda\thanks{Cass Business School, City University,
106 Bunhill Row, UK - London EC1Y 8TZ}},
 \\
\textbf{Jens Perch Nielsen$^\dag$}
 \textbf{ and Stefan Sperlich\thanks{Université de Genève, Département des sciences économiques,
Bd du Pont d'Arve 40, CH - 1211 Genève 4}} \\
}

\maketitle

\begin{abstract}

Recent contributions to kernel smoothing show that the performance of cross-validated
bandwidth selectors improve significantly from indirectness. Indirect crossvalidation
first estimates the classical cross-validated bandwidth from a more rough and difficult
smoothing problem than the original one and then rescales this indirect bandwidth to
become a bandwidth of the original problem. The motivation for this approach comes from the observation that classical
crossvalidation tends to work better when the smoothing problem is difficult. In this paper we
find that the performance of indirect crossvalidation improves theoretically and practically
when the polynomial order of the indirect kernel increases, with the Gaussian kernel as limiting kernel when the polynomial order goes to infinity.
These theoretical and practical results support the often proposed choice of the Gaussian kernel as
 indirect kernel. However, for do-validation our study shows a discrepancy between asymptotic theory and practical performance. As for indirect crossvalidation, in asymptotic theory the performance of indirect do-validation improves with increasing polynomial order of the used indirect kernel. But this theoretical improvements do not carry over to practice and the original do-validation still
seems to be our preferred bandwidth selector. We also consider plug-in estimation and combinations of plug-in bandwidths and crossvalidated bandwidths. These latter bandwidths do not outperform the original do-validation estimator either.
\end{abstract}
\newpage


\newpage

\section{Introduction}

\label{sec:int}

This paper is a study on some theoretical and practical findings on recent proposals for crossvalidated bandwidth selection in kernel density estimation.
Indirect crossvalidation has recently been considered in Hart and Yi (1998),
Hart and Lee (2005), Savchuk, Hart and Sheather (2010a,b).  In this approach in a first step the classical cross-validated bandwidth is calculated for another choice of "indirect" kernel.  The indirect kernel is chosen such that the smoothing problem becomes much more rough and difficult
 than the original one. In a second step the bandwidth is rescaled to
become efficient for the original kernel. The motivation for this approach comes from the empirical finding that classical
crossvalidation tends to work better when the smoothing problem is difficult. In this paper we will compare this approach for polynomial choices of indirect kernels and for Gaussian kernels. In the above papers it was proposed to
use Gaussian kernels in the indirect step. In this paper we give theoretical and practical evidence into this choice. Indeed, as we will see, the Gaussian kernel can be considered as limit of polynomial kernels and it is theoretically optimal in this class of kernels. It is very interesting and comforting to notice that when it comes to indirect crossvalidation than theoretical and practical results go hand in hand. The higher the polynomial order of the indirect kernel, the better the indirect crossvalidation performs in practice as well as in theory.

It turns out that a similar theoretical result applies for Do-validation: again we get theoretical improvements by increasing the order of polynomial indirect kernels. But simulations do not support these findings. Do-validation is an indirect crossvalidation approach that was introduced in Mammen et al.\ (2011). They concluded that do-validation is the indirect crossvalidation method for densities with the best ISE
performance in practice so far. They also argued that when the asymptotic theory of the bandwidth selector gets so good as in the do-validation case, then theoretical improvements of bandwidth selectors are not really important. When theoretical properties become so good it is the practicalities around the implementation that count and here do-validation is excellent. This paper is yet another support of this conclusion on do-validation. While increasing the polynomial order of the indirect kernel improves the practical performance of indirect crossvalidation, this  is no longer true for indirect do-validation. An overall judgement of performance  favors the original do-validation procedure even though asymptotic theory suggests something else. From our simulated results we can see that the problem seems to be that the bias of the bandwidth selector increases with the order of the indirect kernel. This increase in bias is compensated by the decrease of volatility in the indirect crossvalidation case, but not in the indirect do-validation case. This type of conclusion is not foreign to us, because it parallels our experience with plug-in methods. Theoretically plug-in methods are  better than indirect do-validation with the Gaussian kernel, even though it is a close race between the two. However, plug-in suffers from high bias and it is beat badly in all our finite sample performance measures. We also tried to combine plug-in, indirect crossvalidation and indirect do-validation. We present one such very successful combination. However, the performance of this combination is very similar to the performance of do-validation leaving us to prefer the do-validation because of its simpler computational properties. When we use the 90 percent integrated squared error as our performance measure instead of the classical mean integrated  square error, then the advantage of do-validation and our new median estimator  is even more significant. Also, the 90 percent quantile might  be closer to what applied statisticians are looking for when evaluating the performance of a bandwidth selector. Applied statisticians might not be so impressed of their estimated nonparametric curve behaving well in some average sense. They are perhaps more interested in whether the concrete curve they have in front of them is as good as it can get. Looking at the quantile where you are worse 10 percent of the time and better 90 percent of the time give that kind of reassurance. In particular when the results are as crystal clear as the finite sample results we are getting in this  paper. The bandwidth selectors of this paper could potentially also carry over to other smoothing problems, see Soni, Dewan and Jain (2012) and Oliveira, Crujeiras and Rodríguez-Casal (2012) and G\'amiz-P\'erez, Mart{\'\i}nez-Miranda and Nielsen (2012). The latter paper actually supports the use of Do-validation in survival smoothing and show its superiority to classical crossvalidation also in this case.  The new insights provided by this current paper on Do-validation add to the confidence that Do-validation will be useful even beyond the simplest possible iid setting considered here.

The paper is organized as follows. In Section \ref{sec:icv} we
first consider indirect crossvalidation, where the theoretical and practical
improvements of highering the order of the indirect kernel is very clear.
Both the theoretical and the finite sample performance improve consistently in
every step we increase the order of the indirect kernel. In
Section \ref{sec:ido} we consider indirect do-validation. The theoretical
relative improvements of highering the order of the indirect kernel are very similar
to those we saw for classical crossvalidation. However, the finite sample results are
less clear. Here increasing the order of the indirect kernel often helps, but not always.
But as for indirect crossvalidation, we find very different bias/variance trade-offs
for the different  indirect kernels. This motivates the study of kernels that are combinations of several indirect crossvalidation  selectors. In Section \ref{sec:pi} we consider a new and stable median estimator and compare it with plug-in bandwidths and indirect crossvalidation.
Along the next sections we describe simulation experiments to assess the finite sample performance of all the methods. The simulations scenario is described for all the cases in Subsection \ref{subsec:sim}.

\section{Indirect cross-validated bandwidth selection in kernel density estimation}
\label{sec:icv}

In this section we consider indirect crossvalidation in its simplest possible version taken from
Hart and Lee (2005), Savchuk, Hart and Sheather (2010a,b). These three papers also considered a number of variations of indirect crossvalidation, but we consider in this section the simplest possible version, where one has one indirect kernel and one original kernel. In this section we do not consider one-sided kernels. The above three papers seem to have some preference for the Gaussian kernel as indirect kernel. In this section we are able to give a theoretical justification for why this might be a good idea. The Gaussian kernel is in some sense the optimal kernel of a class of indirect kernels. And the theoretical and practical advantage of choosing the Gaussian kernel as indirect kernel can be quite substantial. In our derivation of indirect crossvalidation below we borrow notation from
Mammen et al.\ (2011) that considered a class of
bandwidth selectors which contains the indirect crossvalidation bandwidth $\widehat{h}_{ICV,L}$ with indirect kernel  $L$ as special case.

The aim is to get a
bandwidth with a small Integrated Squared Error (ISE) for the kernel density
estimator
\begin{equation*}
\widehat{f}_{h,K}(x)=\frac{1}{nh}\sum_{i=1}^{n}K\left( \frac{X_{i}-x}{h}%
\right) .
\end{equation*}%

The bandwidth $\widehat{h}_{ICV,L}$ is based on the inspection of the kernel
density estimator $\widehat{f}_{h,L}$, for a kernel $L$ that fulfills $L(0)=0$. And it comes from the following CV score:
\begin{equation}
\int \widehat{f}%
_{h,L}(x)^{2}dx-2n^{-1}\sum_{i=1}^{n}\widehat{f}_{h,L}(X_{i}).
\label{cross}
\end{equation}%
Note that because of $L(0)=0$ we do not need to use a leave-one-out
version of $\widehat{f}_{h,L}$ in the sum on the right hand side.
The indirect crossvalidation bandwidth $\widehat{h}_{ICV,L}$ is defined by
\begin{equation}
\widehat{h}_{ICV,L}=\left( \frac{R(K)}{\mu _{2}^{2}(K)}\frac{\mu_{2}^{2}(L)}{R(L)}\right) ^{1/5}\widehat{h}_{L}
\label{def-icv}
\end{equation}%
with $\widehat{h}_{L}$ being the minimizer of the score (\ref{cross}). Here $R(g)=\int g^{2}(x)dx$, $\mu _{l}(g)=\int x^{l}g(x)dx$ for functions $%
g $ and integers $l\geq 0$. Note that the bandwidth $\widehat{h}_{L}$ is a selector
for the density estimator with kernel $L$. After multiplying with the factor $(R(K)\mu
_{2}^{2}(L))^{1/5}({\mu _{2}^{2}(K)}$ ${R(L)})^{-1/5}$ it becomes a
selector for the density estimator $\widehat{f}_{h,K}$. This follows from classical smoothing theory
and has been used at many places in the discussion of bandwidth selectors.
Note that the indirect crossvalidation method contains the classical crossvalidation
bandwidth selector as one example with $K=L$.

We now apply results from Mammen et al.\ (2011) to derive
 the asymptotic distribution of the difference between the indirect crossvalidation bandwidths $\widehat h_{ICV,L}$ and the ISE-optimal bandwidth $h_{ISE}$. Here, the bandwidth $h_{ISE}$ is defined by
\begin{eqnarray*}
h_{ISE} = \mbox{arg} \min_h \left [ \int \left (\widehat f_{h,K}(x)- f(x)
\right )^2 dx \right ].
\end{eqnarray*}
Under some mild conditions on the density $f$ and the kernels $K$ and  $L$, see Assumptions (A1) and (A2) in Mammen et al.\ (2011),  one gets by application of their Theorem 1 that for symmetric kernels $K$ and $L$ \begin{eqnarray}
n^{3/10}(\widehat{h}_{ICV,L}-h_{ISE})\rightarrow N(0,\sigma _{L,ICV}^{2}) &&%
\mbox{in
distribution},  \label{asympeqA}
\end{eqnarray}%
where
\begin{equation}
\begin{array}{rl}
\sigma _{L,ICV}^{2}= & \displaystyle\frac{4}{25}R(K)^{-2/5}\mu _{2}^{-6/5}(K)R({%
f^{\prime \prime }})^{-8/5}\mathop{\rm V}\nolimits(f^{\prime \prime })+\displaystyle\frac{1}{50}R(K)^{-7/5}\mu _{2}^{-6/5}(K) \\
& \ \ \times R({f^{\prime \prime }})^{-3/5}R(f)\displaystyle\int \left[
H(u)-\left( \displaystyle\frac{%
R(K)}{R(L)}\right) H_{ICV,L}(u)\right] ^{2}du,%
\end{array}
\label{asympvar}
\end{equation}%
with
\begin{eqnarray*}
\mathop{\rm V}\nolimits(f^{\prime \prime }) &=&\int f^{\prime \prime
2}(x)f(x)dx-\left( \int f^{\prime \prime }(x)f(x)dx\right) ^{2}, \\
&& \\
H(u) &=&4\int K(u-v)[K(v)+vK^{\prime }(v)]dv, \\
&& \\
H_{ICV,L}(d_{L}u) &=&4\int L(u-v)[L(v)+vL^{\prime
}(v)]dv -4\left[ L(u)+uL^{\prime }(u)\right]
, \\
&& \\
d_{L} &=&\left( \frac{R(K)}{R(L)}\frac{\mu _{2}^{2}(L)}{\mu
_{2}^{2}(K)}\right) ^{-1/5}.\end{eqnarray*}

Here we are interested in indirect crossvalidation defined with any symmetric kernel $K$ (as for example the Epanechnikov kernel) and as the kernel $L$ a polynomial kernel with higher order. Specifically we define a general kernel function with order $r$ by
\begin{equation} \label{def:K2r}
K_{2r}(u)= \kappa_r(1-u^2)^r 1_{\{-1<u<1\}}
\end{equation}
 with $\kappa_r = (\int_{-1}^1 (1-u^2)^r du)^{-1}$. Note that for $r=1$ it is the Epanechnikov kernel and for $r=2$ it gives the quartic kernel. We now study the theoretical performance of indirect crossvalidation for the choice $K=K_2$ and $L=K_{2r}$ for different choices of $r$. We start by considering the limiting case $r\to \infty$. For this purpose we consider the kernel $K^*_{2r}(u)= (2r)^{-1/2} K_{2r}((2r)^{-1/2} u )$ that differs from $K_{2r}$ by scale. Because the definition of the bandwidth selector does not depend on the scale of $L$ we have that $\sigma _{K_{2r},ICV}^{2} = \sigma _{K^*_{2r},ICV}^{2}$. Furthermore, because of $\lim_{r \to \infty} (1-(2r)^{-1}u^2)^r = e^{-u^2 /2}$ it holds that, after scaling,
 the polynomial kernels converge to the  Gaussian kernel  when $r$ goes to infinity
 $$
  \lim_{r\to \infty} (2r)^{-1/2} K_{2r}((2r)^{-1/2} u) = \phi(u)= \frac{1}{\sqrt{2\pi}} e^{-\frac{u^2}{2}}.
  $$
  Moreover, it holds that  $\sigma _{K_{2r},ICV}^{2} = \sigma _{K^*_{2r},ICV}^{2} \to \sigma _{\phi,ICV}^{2}$ for $r \to \infty$. This can be shown by dominated convergence
 using the fact that $(1-(2r)^{-1}u^2)^r \leq e^{-u^2 /2}$. Thus a Gaussian indirect kernel is a limiting case for the performance of indirect crossvalidation.

\begin{figure}[htb]
\begin{center}
 \includegraphics[width=11cm]{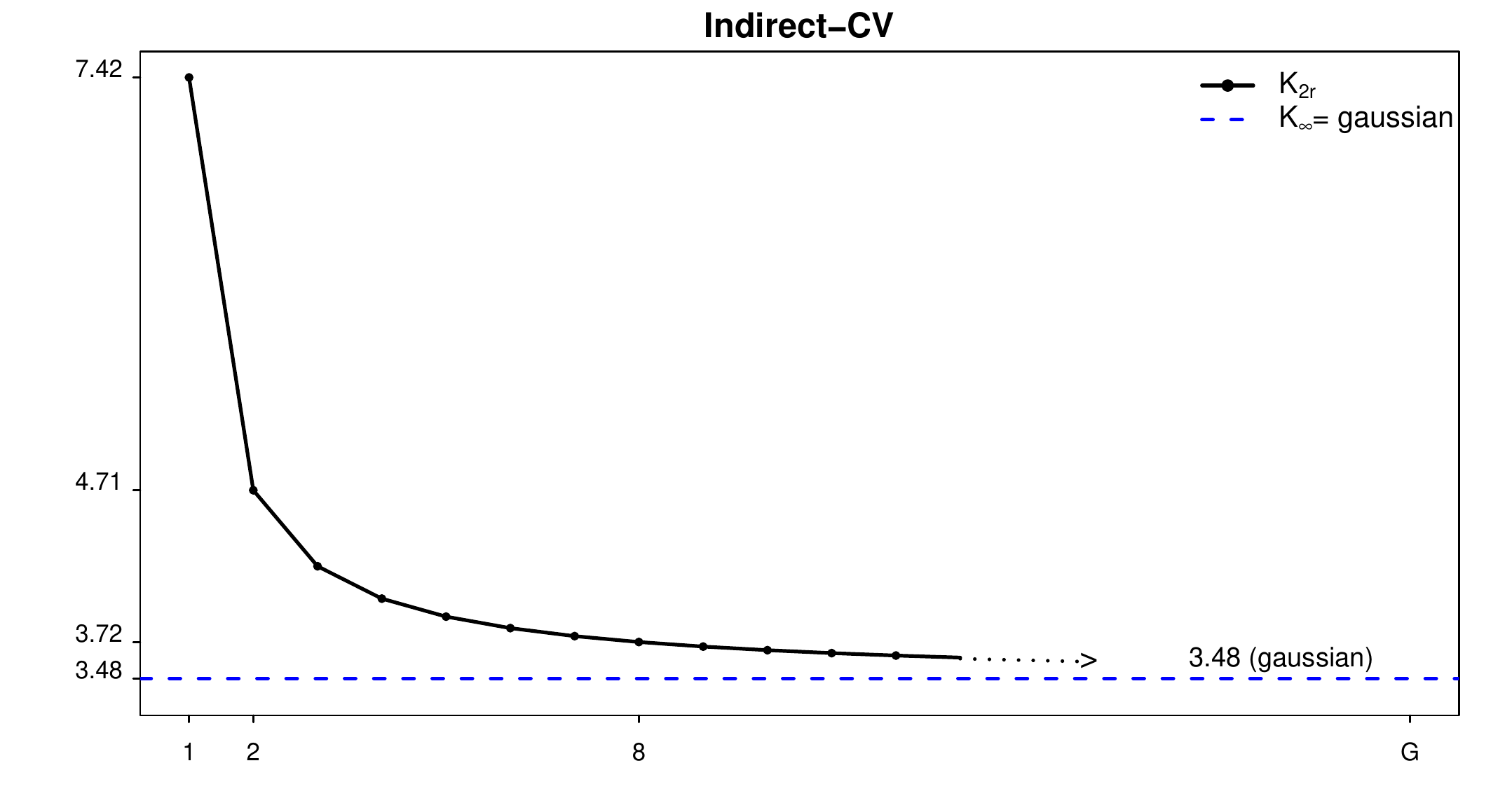}
\end{center}
\vspace{-1cm}
\caption{Asymptotic variance reduction for indirect crossvalidation with kernels $K_{2r}$ for $r=1,2,\ldots$ and the Gaussian limit.\label{Fig:K2r-icv}}
\end{figure}

 According to (\ref{asympeqA}) and (\ref {asympvar}), the asymptotic variance of $\widehat{h}_{ICV,K_{2r}}-h_{ISE}$ is of the form $C_{f,K}\left\{ 4R(K)\frac{
\mathop{\rm V}\nolimits(f^{\prime \prime })}{R({f^{\prime \prime }})R(f)}+
c_r \right \}$ with a constant $c_r$ depending on $r$ and with $C_{f,K}$ as a function of $f$ and $K$. We have just argued that $c_r \to c_{\infty}$ for $r \to \infty$ where $c_{\infty}=3.48$ is the constant corresponding to the Gaussian kernel. Figure \ref{Fig:K2r-icv} shows $c_r$ as a function of $r$. It illustrates the convergence but it also shows that this convergence is monotone: by increasing the order $r$ ($r=2,3,4,\ldots$) we get an incremental reduction in the asymptotic variance factor for indirect crossvalidation.

One sees that the trick of indirect
crossvalidation significantly improves on crossvalidation. And specifically the asymptotics for the indirect crossvalidatory bandwidths with kernels $K$ and $L$ being  the Epanechnikov kernel and $L=K_{2r}$, respectively are given below for $r=1,2,8$ and $r \to \infty$, which becomes the Gaussian kernel.
\begin{eqnarray*}
\sigma _{\mathop{\rm CV}\nolimits}^{2} &=&C_{f,K}\left\{ 4R(K)\frac{%
\mathop{\rm V}\nolimits(f^{\prime \prime })}{R({f^{\prime \prime }})R(f)}+{%
7.42}\right\}  \\
\sigma _{\mathop{\rm ICV}_2\nolimits}^{2} &=&C_{f,K}\left\{ 4R(K)\frac{%
\mathop{\rm V}\nolimits(f^{\prime \prime })}{R({f^{\prime \prime }})R(f)}+{%
4.71}\right\}  \\
\sigma _{\mathop{\rm ICV}_8\nolimits}^{2} &=&C_{f,K}\left\{ 4R(K)\frac{%
\mathop{\rm V}\nolimits(f^{\prime \prime })}{R({f^{\prime \prime }})R(f)}+{%
3.72}\right\}  \\
\sigma _{\mathop{\rm ICV}_G\nolimits}^{2} &=&C_{f,K}\left\{ 4R(K)\frac{%
\mathop{\rm V}\nolimits(f^{\prime \prime })}{R({f^{\prime \prime }})R(f)}+{%
3.48}\right\} . \\
\end{eqnarray*}%
 The first improvement of going from standard crossvalidation to having an indirect kernel of one more order is the most important one. The crucial component of the asymptotic theory is decreasing from 7.42 to 4.71. This is sufficiently substantial to consider this simple adjustment of classical crossvalidation to solving a good and important part of the problem with the volatility of the crossvalidation estimator. However, indirect crossvalidation can do better. Going to the Gaussian limit brings the crucial constant down to 3.48! This is quite low and approaching the do-validation constant of 2.19 found in Mammen et al.\ (2011). It turns out that 3.48 is still so big that 2.19 is a major improvement in theory and practice. Do-validation does better than indirect crossvalidation in theory and practice, even when the latter is based on the optimal Gaussian kernel.

\subsection{Simulation experiments about indirect crossvalidation}
\label{subsec:sim}

\renewcommand{\baselinestretch}{1.2}
\begin{table}[hbt]
\begin{center}
{\small {\tabcolsep4.5pt
\begin{tabular}{c|rrrrr|rrrrr}
\multicolumn{1}{c}{} & \multicolumn{5}{c}{Design 1} & \multicolumn{5}{|c}{
Design 2} \\ \hline
\multicolumn{1}{c}{}  & {$h_{ISE}$}
& {\scriptsize {$\widehat h_{CV}$}} & {\scriptsize {$\widehat h_{ICV_2}$}} &
{\scriptsize {$\widehat h_{ICV_8}$}} & {\scriptsize {$\widehat h_{ICV_G}$}}
& {$h_{ISE}$}
& {\scriptsize {$\widehat h_{CV}$}} & {\scriptsize {$\widehat h_{ICV_2}$}} &
{\scriptsize {$\widehat h_{ICV_8}$}} & {\scriptsize {$\widehat h_{ICV_G}$}} \\
\hline
\multicolumn{1}{c}{}  & \multicolumn{10}{c}{$n=100$} \\ \hline
$m_1$ & 2.328 &4.944 & 4.804 & 4.583 & 4.446 & 3.477 & 6.313 & 6.047 & 5.876 & 5.809\\
$m_2$ & 1.876 &6.185 & 5.557 & 5.451 & 5.256 & 1.989 & 5.611 & 5.089 & 4.683 & 4.511\\
$m_3$ & 0.000 &4.963 & 4.087 & 3.840 & 3.472 & 0.000& 2.550 &1.969 &1.969 &1.969\\
$m_4$ & 0.000 &-1.839 &-1.044 &-0.458 & -0.230& 0.000& 0.587 & 1.049 & 1.532 & 1.746\\
$m_5$ & 0.000 &0.714  & 0.724 &0.724 &0.722 & 0.000 & 0.943 &0.948 &0.950& 1.000\\ \hline
\multicolumn{1}{c}{}  & \multicolumn{10}{c}{$n=200$} \\ \hline
$m_1$ & 1.417 & 2.573 & 2.481 & 2.359 & 2.288 & 2.307 & 3.816 & 3.700 & 3.495 & 3.451\\
$m_2$ & 1.098  &2.747 & 2.609 & 2.435 & 2.213 & 1.372 & 3.376 & 3.239 & 2.577 & 2.533\\
$m_3$ & 0.000 &3.161 &2.545 &2.133 &   1.989  & 0.000 & 2.174 & 1.821 & 1.487 & 1.477\\
$m_4$ & 0.000 &-0.718 &-0.438 & 0.024 & 0.192 & 0.000 &-0.087 & 0.240 & 0.593 & 0.769\\
$m_5$ & 0.000 &0.687 & 0.690 &0.687 &0.687& 0.000 & 0.723 &0.751 &0.733 &0.737\\ \hline
\multicolumn{1}{c}{}  & \multicolumn{10}{c}{$n=500$} \\ \hline
$m_1$ & 0.731  &1.221 & 1.175 & 1.129 & 1.108 & 1.208 & 1.780 & 1.756 & 1.695 & 1.674\\
$m_2$ & 0.465  &1.078 & 1.027 & 0.913 & 0.867 & 0.648 & 1.237 & 1.245 & 1.147 & 1.122\\
$m_3$ & 0.000 &  2.615 & 2.214 & 1.935 & 1.818& 0.000&  1.296 & 1.218 & 1.076 & 0.997\\
$m_4$ & 0.000 &-0.805 &-0.417 &-0.193 &-0.104 & 0.000 &-0.195 & 0.008 & 0.193 & 0.285\\
$m_5$ & 0.000 &0.666 & 0.666 & 0.651 & 0.656& 0.000 & 0.632 & 0.629& 0.632& 0.634 \\ \hline
\multicolumn{1}{c}{}  & \multicolumn{10}{c}{$n=1000$} \\ \hline
$m_1$ &0.439  & 0.719 & 0.712 & 0.675 & 0.664 & 0.732 & 1.049 & 1.006 & 0.987 & 0.976 \\
$m_2$ &0.277  & 0.699 & 0.699 & 0.622 & 0.606 & 0.377 & 0.722 & 0.624 & 0.609 & 0.599 \\
$m_3$ & 0.000 & 2.190 & 2.161 & 1.741 & 1.688& 0.000&  1.227 & 1.071 &0.914 &0.857\\
$m_4$ &0.000  &-0.596 &-0.434 &-0.236 &-0.155 & 0.000 &-0.201 &-0.074 & 0.051 & 0.119 \\
$m_5$ & 0.000 &0.667 & 0.667 & 0.643 & 0.632& 0.000 & 0.586 &  0.560 & 0.554 & 0.554\\ \hline
\end{tabular}
}}
\end{center}
\caption{\textit{Simulation results about the indirect crossvalidation method with designs 1 and 2. We compare the standard crossvalidation, $\widehat h_{CV}$, with three indirect versions $\widehat h_{ICV_2}$, $\widehat h_{ICV_8}$ and $\widehat h_{ICV_G}$ for kernels $K_{2r}$ with $r=2,8, \infty$. As a benchmark we report the results for the unfeasible ISE optimal bandwidth, $h_{ISE}$.}}
\label{tab-ICV-1}
\end{table}

\renewcommand{\baselinestretch}{1.2}
\begin{table}[hbt]
\begin{center}
{\small {\tabcolsep4.5pt
\begin{tabular}{c|rrrrr|rrrrr}
\multicolumn{1}{c}{} & \multicolumn{5}{c}{Design 3} & \multicolumn{5}{|c}{
Design 4} \\ \hline
\multicolumn{1}{c}{}  & {$h_{ISE}$}
& {\scriptsize {$\widehat h_{CV}$}} & {\scriptsize {$\widehat h_{ICV_2}$}} &
{\scriptsize {$\widehat h_{ICV_8}$}} & {\scriptsize {$\widehat h_{ICV_G}$}}
& {$h_{ISE}$}
& {\scriptsize {$\widehat h_{CV}$}} & {\scriptsize {$\widehat h_{ICV_2}$}} &
{\scriptsize {$\widehat h_{ICV_8}$}} & {\scriptsize {$\widehat h_{ICV_G}$}} \\
\hline
\multicolumn{1}{c}{}  & \multicolumn{10}{c}{$n=100$} \\ \hline
$m_1$ &4.448  &7.232  &7.061  &6.905  &6.951  &4.842  & 7.918 & 7.818 & 7.636 &7.643  \\
$m_2$ &2.231  &6.392  &6.141  &5.326  &5.247  &2.644  & 6.698 & 6.842 & 6.400 &6.440  \\
$m_3$ & 0.000 & 1.766 &1.515 &1.461 & 1.526  & 0.000  & 1.595 &1.460 &1.357 &1.328\\
$m_4$ &0.000  &0.629  &1.008  &1.423  &1.705  &0.000  &-0.256 & 1.146 & 0.569 &0.742  \\
$m_5$ & 0.000 & 0.824 &0.883 &0.957 &1.060& 0.000 &  0.822 &0.842& 0.844 &0.869 \\ \hline
\multicolumn{1}{c}{}  & \multicolumn{10}{c}{$n=200$} \\ \hline
$m_1$ &2.830  &4.216  &4.034  &3.872  &3.864  &3.100  &4.643  &4.521  &4.453  &4.405  \\
$m_2$ &1.343  &3.043  &2.788  &2.310  &2.299  &1.657  &3.645  &3.299  &3.391  &3.265  \\
$m_3$ & 0.000 & 1.244 &1.016 &0.947 &0.932   & 0.000& 1.396 & 1.228 & 1.106 & 1.118\\
$m_4$ &0.000  &0.086  &0.291  &0.593  &0.707  &0.000  &-0.313 &-0.042 &0.233  &0.360  \\
$m_5$ & 0.000 & 0.626 &0.649 &0.626 &0.648& 0.000 &  0.758 & 0.765 &0.765& 0.778  \\ \hline
\multicolumn{1}{c}{}  & \multicolumn{10}{c}{$n=500$} \\ \hline
$m_1$ &1.540  &2.006  &1.955  &1.908  &1.889  &1.687  &2.338  &2.270  &2.193  &2.164  \\
$m_2$ &0.685  &1.053  &0.998  &0.994  &0.963  &0.767  &1.576  &1.516  &1.333  &1.272  \\
$m_3$ & 0.000 & 0.859 &0.812 &0.673  &0.640   & 0.000& 0.924 & 0.826 & 0.721 & 0.717\\
$m_4$ &0.000  &-0.074 &0.033  &0.187  &0.271  &0.000  &-0.436 &-0.205 &-0.031 &0.073  \\
$m_5$ & 0.000 &  0.562 &0.532 &0.532 &0.532& 0.000 &  0.625 & 0.611 & 0.587 &0.587 \\ \hline
\multicolumn{1}{c}{}  & \multicolumn{10}{c}{$n=1000$} \\ \hline
$m_1$ & 0.943 & 1.166 & 1.135 & 1.112 & 1.109 &1.060 &1.341  &1.317  &1.281  &1.270  \\
$m_2$ & 0.405 & 0.620 & 0.590 & 0.533 & 0.536 &0.491 &0.725  &0.706  &0.645  &0.632   \\
$m_3$ & 0.000 & 0.683 &0.553 & 0.457 & 0.444 & 0.000& 0.784 & 0.639 & 0.549 & 0.504 \\
$m_4$ & 0.000 & -0.118& 0.003 & 0.088 & 0.139 &0.000 &-0.287 &-0.132 &0.030  &0.114  \\
$m_5$ & 0.000 &  0.446 &0.467& 0.428 &0.438 & 0.000 &   0.564 & 0.538& 0.528& 0.528  \\ \hline
\end{tabular}
}}
\end{center}
\caption{\textit{Simulation results about the indirect crossvalidation method with designs 3 and 4. We compare the standard crossvalidation, $\widehat h_{CV}$, with three indirect versions $\widehat h_{ICV_2}$, $\widehat h_{ICV_8}$ and $\widehat h_{ICV_G}$ for kernels $K_{2r}$ with $r=2,8, \infty$. As a benchmark we report the results for the unfeasible ISE optimal bandwidth, $h_{ISE}$.}}
\label{tab-ICV-2}
\end{table}

\renewcommand{\baselinestretch}{1.2}
\begin{table}[hbt]
\begin{center}
{\small {\tabcolsep4.5pt
\begin{tabular}{c|rrrrr|rrrrr}
\multicolumn{1}{c}{} & \multicolumn{5}{c}{Design 5} & \multicolumn{5}{|c}{
Design 6} \\ \hline
\multicolumn{1}{c}{}  & {$h_{ISE}$}
& {\scriptsize {$\widehat h_{CV}$}} & {\scriptsize {$\widehat h_{ICV_2}$}} &
{\scriptsize {$\widehat h_{ICV_8}$}} & {\scriptsize {$\widehat h_{ICV_G}$}}
& {$h_{ISE}$}
& {\scriptsize {$\widehat h_{CV}$}} & {\scriptsize {$\widehat h_{ICV_2}$}} &
{\scriptsize {$\widehat h_{ICV_8}$}} & {\scriptsize {$\widehat h_{ICV_G}$}} \\
\hline
\multicolumn{1}{c}{}  & \multicolumn{10}{c}{$n=100$} \\ \hline
$m_1$ & 3.356 & 5.575 & 5.488 & 5.250 & 5.208 & 3.633 & 5.458 & 5.279 & 5.184 & 5.149 \\
$m_2$ & 1.383 & 6.160 & 6.065 & 4.638 & 4.575 & 1.617 & 4.309 & 3.829 & 3.245 & 3.140 \\
$m_3$ & 0.000 & 1.730 & 1.624 & 1.458 & 1.380 & 0.000 & 1.332 & 1.121 & 1.122 & 1.122 \\
$m_4$ & 0.000 &-0.101 & 0.616 & 1.270 & 1.521 & 0.000 & 0.749 & 1.369 & 1.970 & 2.279  \\
$m_5$ & 0.000 &  0.920 &0.999 &0.999 &0.999 & 0.000 & 0.917 & 0.950 & 0.999 & 0.999 \\ \hline
\multicolumn{1}{c}{}  & \multicolumn{10}{c}{$n=200$} \\ \hline
$m_1$ &2.293  &3.516  &3.400  &3.269  &3.223  &2.387  &3.397  &3.317  &3.220  &3.212  \\
$m_2$ &0.864  &2.907  &2.483  &2.309  &2.194  &0.955  &2.378  &2.209  &2.014  &1.949  \\
$m_3$ & 0.000 & 1.551 &1.425 & 1.289  &1.248 & 0.000 & 1.160 &1.040 &0.965 &0.962 \\
$m_4$ &0.000  &-0.370 &0.158  &0.638  &0.833  &0.000  &0.373  &0.672  &1.020  &1.194  \\
$m_5$ & 0.000 &  0.791 &0.819 &0.825 &0.807 & 0.000 &  0.856 &0.839 &0.839 &0.838  \\ \hline
\multicolumn{1}{c}{}  & \multicolumn{10}{c}{$n=500$} \\ \hline
$m_1$ &1.287  &1.857  &1.806  &1.758  &1.729  &1.355  &1.823  &1.746  &1.719  &1.700  \\
$m_2$ &0.520  &1.329  &1.238  &1.176  &1.081  &0.495  &1.150  &0.889  &0.885  &0.821  \\
$m_3$ & 0.000 & 1.298 & 1.104 &0.964  &0.930 & 0.000&  0.973 & 0.886 &0.872 &0.788 \\
$m_4$ &0.000  &-0.093 &0.143  &0.439  &0.602  &0.000  &-0.333 &-0.045 &0.244  &0.399  \\
$m_5$ & 0.000 & 0.760 & 0.774 &0.751 & 0.762 & 0.000 &  0.637 & 0.621 &0.618 &0.618  \\ \hline
\multicolumn{1}{c}{}  & \multicolumn{10}{c}{$n=1000$} \\ \hline
$m_1$ & 0.844 & 1.147 & 1.102 & 1.075 & 1.067 & 0.892 & 1.074 & 1.054 & 1.032 & 1.024 \\
$m_2$ & 0.357 & 0.691 & 0.611 & 0.565 & 0.546 & 0.304 & 0.458 & 0.445 & 0.393 & 0.370 \\
$m_3$ & 0.000 & 1.053 & 0.866 & 0.802 &0.729 & 0.000&  0.500 &0.465 &0.403 &0.441 \\
$m_4$ & 0.000 & -0.300& -0.109& 0.133 & 0.245 & 0.000 & -0.149& 0.049 & 0.253 & 0.381 \\
$m_5$ & 0.000 &  0.667 & 0.600 &0.586& 0.591 & 0.000 &   0.500& 0.516& 0.520& 0.500 \\ \hline
\end{tabular}
}}
\end{center}
\caption{\textit{Simulation results about the indirect crossvalidation method with designs 5 and 6. We compare the standard crossvalidation, $\widehat h_{CV}$, with three indirect versions $\widehat h_{ICV_2}$, $\widehat h_{ICV_8}$ and $\widehat h_{ICV_G}$ for kernels $K_{2r}$ with $r=2,8, \infty$. As a benchmark we report the results for the unfeasible ISE optimal bandwidth, $h_{ISE}$.}}
\label{tab-ICV-3}
\end{table}

The purpose of this section is to study the performance of the indirect crossvalidation method with respect to standard crossvalidation and the optimal ISE bandwidth ($h_{ISE}$). We consider in the study three possible indirect crossvalidatory bandwidths: $\widehat h_{ICV_2}$, $\widehat h_{ICV_8}$ and $\widehat h_{ICV_G}$, which comes from using as the kernel $K$ the Epanechnikov kernel and as kernel $L$ the higher order polynomial kernel, $K_{2r}$ defined in (\ref{def:K2r}), for $r=2,8, \ldots,\infty$ with $K_{\infty}$ being the Gaussian kernel.

\begin{figure}[htb]
\begin{center}
 \vspace{-11.0cm}
 \includegraphics[width=16.5cm,height=16.5cm]{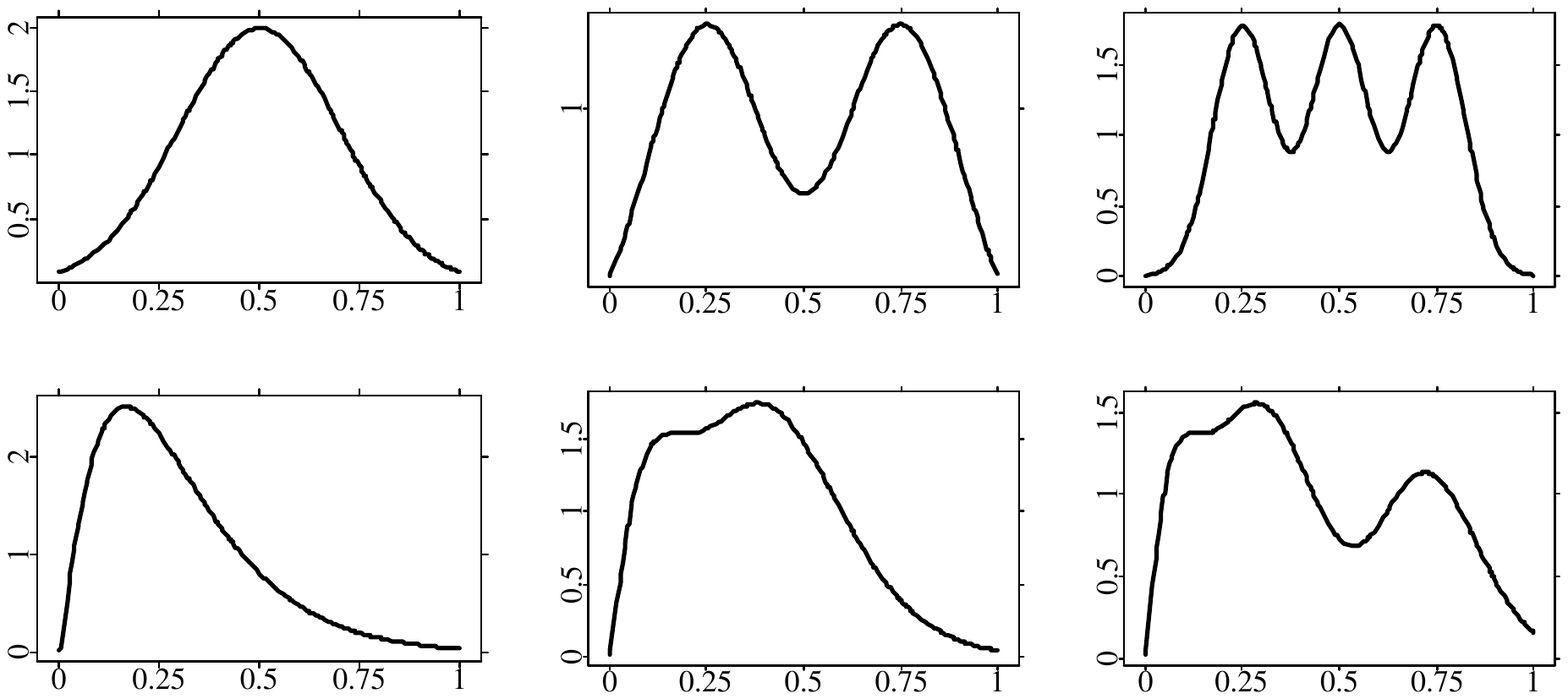}
\end{center}
 \vspace{-1.1cm}
\caption{The six data generating densities: Designs 1 to 6 from the
upper left to the lower right.\label{fig-dgp}}
\end{figure}

We consider the same data generating processes as Mammen et al.\ (2011). We simulate six designs defined by the six densities plotted in
Figure \ref{fig-dgp} and defined as follows:

\begin{itemize}
\item[1.] a simple normal distribution, $N(0.5,0.2^2)$,

\item[2.] a bimodal mixture of two normals which were $N(0.35, 0.1^2) $ and $%
N(0.65, 0.1^2)$,

\item[3.] a mixture of three normals, namely $N(0.25, 0.075^2) $, $%
N(0.5,0.075^2)$ and $N(0.75, 0.075^2)$ giving three clear modes,

\item[4.] a gamma distribution, $Gamma (a,b)$ with $b=1.5$, $a=b^2$ applied
on $5x$ with $x\in\R_+$, i.e.
\begin{equation*}
f(x) = 5 \frac{b^a}{\Gamma (a)} (5x)^{a-1} e^{-5xb},
\end{equation*}

\item[5.] a mixture of two gamma distributions, $Gamma (a_j,b_j)$, $j=1,2$
with $a_j=b_j^2$, $b_1=1.5$, $b_2=3$ applied on $6x$, i.e.
\begin{equation*}
f(x) = \frac{6}{2} \sum_{j=1}^2 \frac{b_j^{a_j}}{\Gamma (a_j)} (6x)^{a_j-1}
e^{-6xb_j}
\end{equation*}
giving one mode and a plateau,

\item[6.] and a mixture of three gamma distributions, $Gamma (a_j,b_j)$, $%
j=1,2,3$ with $a_j=b_j^2$, $b_1=1.5$, $b_2=3$, and $b_3=6$ applied on $8x$
giving two bumps and one plateau.
\end{itemize}

Our set of densities contains density functions with one, two or three
modes, some being asymmetric. They all have exponentially falling tails,
because otherwise one has to work with boundary correcting kernels. The main
mass is always in $[0,1]$. For the purposes of this paper we use five measures to summarize the
stochastic performance of any bandwidth selectors $\widehat h$:
\begin{align}
m_1&=\mathop{\rm mean}\nolimits({\mathop{\rm ISE}\nolimits}(\widehat h)) \label{m1}  \\
m_2&=\mathop{\rm std}\nolimits({\mathop{\rm ISE}\nolimits}(\widehat h))  \label{m2}\\
m_3& = 90\% quantile \left( | ISE(\hat h) -  ISE(h_{ISE}) | / ISE(h_{ISE}) \right)  \label{m3}\\
m_4&= \mathop{\rm mean}(\widehat h - h_{ISE})   \label{m4}\\
m_5& = 90\% quantile \left( | \hat h -h_{ISE} | / h_{ISE} \right).  \label{m5}
\end{align}
The above measures have been calculated from 500 simulated samples from each density and four samples sizes $n=100,200,500$ and $1000$. The measures $m_1$, $m_2$ and $m_4$ where also used in the simulations by the former paper by Mammen et al.\ (2011). Here we have included measures $m_3$ and $m_5$ which are informative about the stability of the bandwidth estimates. Tables \ref{tab-ICV-1}, \ref{tab-ICV-2} and \ref{tab-ICV-3} show the simulation results. Note that the bias ($m_4$) is consistently increasing as a function of the order of the indirect kernel with the indirect Gaussian kernel having the largest bias. This increase in bias is being more than balanced by a decreasing volatility ($m_2$) as a function of the order of the indirect kernel. As a result, the overall performance, the averaged integrated squared error performance ($m_1$), is decreasing as a function of the order of the indirect kernel with the Gaussian indirect kernel performing best of all. These results are very clear for all the designs and sample sizes and the indirectness in crossvalidation is indeed working quite well.

\section{Indirect do-validation in kernel density estimation}
\label{sec:ido}

Here we describe the indirect do-validation method and provide theoretical and empirical results in a similar way to that for indirect crossvalidation above. We conclude that indirect do-validation improves consistently theoretically when the
order of the indirect kernel increases. The relative improvements parallel those we saw for
indirect crossvalidation.  However, it does not seem like the practical improvements follow
the theoretical improvements for indirect do-validation. The original conclusion of
Mammen et al.\ (2011) seems to be valid also here:
``when the theoretical properties are so good as in do-validation, it is the practical
implementation at hand that counts, not further theoretical improvements''.
Going all the way to the limiting Gaussian kernel is not of practical relevance for indirect do-validation.

In our derivation of the methodology, we follow Mammen et al.\ (2011) that
first consider a class of
bandwidth selectors that are constructed as weighted averages
of crossvalidation bandwidths. This class of bandwidth selectors contains the classical crossvalidation
bandwidth selector as one example with $J=1$ and $L_{1}(u)=K(u)\mathbf{1}%
(u\not=0)$. And it also contains the do-validation method which combines left and right-sided crossvalidation
using the local linear kernel density estimator (Jones, 1993; and Cheng
1997a, 1997b). In fact the method cannot work on local constant density estimation because of
the inferior rate of convergence it has when applying to asymmetric kernels.
For a kernel density estimator $\widehat{f}_{h,M}$ with kernel $M$ the local
linear kernel density estimator can be defined as kernel density estimator $%
\widehat{f}_{h,M^{\ast }}$ with ``equivalent kernel'' $M^{\ast }$ given by
\begin{equation}
M^{\ast }(u)=\frac{\mu _{2}(M)-\mu _{1}(M)u}{\mu _{0}(M)\mu _{2}(M)-\mu
_{1}^{2}(M)}M(u).  \label{Kequiv}
\end{equation}%
In onesided crossvalidation the basic kernel $M(u)$ is chosen as $2K(u)%
\mathbf{1}_{(-\infty ,0)}$ (leftsided crossvalidation) and $2K(u)\mathbf{1}%
_{(0,\infty )}$ (rightsided crossvalidation). This results in the following
equivalent kernels
\begin{eqnarray}
K_{L}(u) &=&\frac{\mu _{2}(K)+u\mu _{1}^{\ast }(K)}{\mu _{2}(K)-(\mu
_{1}^{\ast }(K))^{2}}2K(u)\mathbf{1}_{(-\infty ,0)},  \label{L2} \\
K_{R}(u) &=&\frac{\mu _{2}(K)-u\mu _{1}^{\ast }(K)}{\mu _{2}(K)-(\mu
_{1}^{\ast }(K))^{2}}2K(u)\mathbf{1}_{(0,\infty )},
\end{eqnarray}%
with $\mu _{1}^{\ast }(K)=2\int_{0}^{\infty }uK(u)du$. Here we have assumed
that the kernel $K$ is symmetric. The left-$\mathop{\rm OSCV}\nolimits$
criterion, denoted by $\mathop{\rm OSCV}\nolimits_{L}$, is defined by
\begin{equation}
\mathop{\rm OSCV}\nolimits_{L}(h)=\int \widehat{f}%
_{h,K_{L}}^{2}(x)dx-2n^{-1}\sum_{i=1}^{n}\widehat{f}_{h,K_{L}}(X_{i}),
\label{oscv}
\end{equation}%
with $\widehat{h}_{L}$ as its minimizer. The left-$\mathop{\rm OSCV}%
\nolimits$ bandwidth is calculated from $\widehat{h}_{L}$ by
\begin{equation} \label{oscv-L}
\widehat{h}_{L,\mathop{\rm OSCV}\nolimits}=\mathrm{C}\widehat{h}_{L},
\end{equation}%
where
\begin{equation}
\mathop{\rm C}\nolimits=\left( \displaystyle\frac{R(K)}{\mu _{2}^{2}(K)}%
\displaystyle\frac{\mu _{2}^{2}(K_{L})}{R(K_{L})}\right) ^{1/5}.  \label{C}
\end{equation}
In exactly the same way we define the right-$\mathop{\rm OSCV}\nolimits$
criterion, $\mathop{\rm OSCV}\nolimits_R$, except that $\widehat f_{h,K_L}$
in (\ref{oscv}) is replaced by $\widehat f_{h,K_R}$. The right-$%
\mathop{\rm
OSCV}\nolimits$ bandwidth is calculated by $\widehat h_{R,\mathop{\rm OSCV}%
\nolimits} = \mathrm{C} \widehat{h}_R$, where $\mathop{\rm
C}\nolimits$ is the same as in (\ref{C}) and $\widehat{h}_R$ is the
minimizer of $\mathop{\rm
OSCV}\nolimits_R$. The do-validation selector $\widehat h_{DO} $ is given by
\begin{equation} \label{h_DO}
\widehat h_{DO} = {\frac{1 }{2}} (\widehat{h}_{L,OSCV} +\widehat{h}_{R,OSCV}
).
\end{equation}
See Mart{\'\i}nez-Miranda, Nielsen and Sperlich (2009) and Mammen et al.\ (2011) for more details.

Left-onesided crossvalidation and right-onesided crossvalidation are not
identical in the local linear case because of differences in the boundary.
However, asymptotically they are equivalent. As we will see in our
simulations do-validation delivers a good stable compromise. It has the same
asymptotic theory as each of the two onesided alternatives and a better
overall finite sample performance.

Again, Theorem 1 in Mammen et al.\ (2011) can be used to get the asymptotic distribution of $\widehat h_{DO} - h_{ISE}$.
Under their Assumptions (A1) and (A2) it holds for symmetric kernel $K$ that
\begin{eqnarray}
n^{3/10}(\widehat{h}_{DO}-h_{ISE})\rightarrow N(0,\sigma_{DO}^{2}) &&%
\mbox{in
distribution},  \label{asympeq1}
\end{eqnarray}%
where
\begin{equation} \label{sigma:do}
\begin{array}{rl}
\sigma_{DO}^{2}= & \displaystyle\frac{4}{25}R(K)^{-2/5}\mu _{2}^{-6/5}(K)R({%
f^{\prime \prime }})^{-8/5}\mathop{\rm V}\nolimits(f^{\prime \prime })+\displaystyle\frac{1}{50}R(K)^{-7/5}\mu _{2}^{-6/5}(K) \\
& \ \ \times R({f^{\prime \prime }})^{-3/5}R(f)\displaystyle\int \left[
H(u)- \displaystyle\frac{%
R(K)}{R(K_{L})} H^*(u)\right] ^{2}du,%
\end{array}
\end{equation}%
with $\mathop{\rm V}\nolimits(f^{\prime \prime })$, $H(u)$ as in the last section and with
\begin{eqnarray*}
H^*(d^*u) &=&2\int K_{L}(u+v)K_{L}(v)dv+2\int K_{L}(-u+v)K_{L}(v)dv \\
&&+2\int K_{L}(u+v)vK_{L}^{\prime }(v)dv+2\int K_{L}(-u+v)vK_{L}^{\prime
}(v)dv \\
&&-2\left[ K_{L}(u)+uK_{L}^{\prime }(u)+K_{L}(-u)-u K_{L}^{\prime }(-u)\right]
, \\
&& \\
d^* &=&\left( \frac{R(K)}{R(K_{L})}\frac{\mu _{2}^{2}(K_{L})}{\mu
_{2}^{2}(K)}\right) ^{-1/5}.\end{eqnarray*}
For the  kernel $K$ equal to the Epanechnikov kernel this gives
\begin{eqnarray*}
\sigma _{\mathop{\rm DO}\nolimits}^{2} &=&C_{f,K}\left\{ 4R(K)\frac{%
\mathop{\rm V}\nolimits(f^{\prime \prime })}{R({f^{\prime \prime }})R(f)}+{%
2.19}\right\}.
\end{eqnarray*}
This can be compared with the asymptotic variance of the Plug-in bandwidth which is equal to
\begin{eqnarray*}
\sigma _{PI}^{2} &=&C_{f,K}\left\{ 4R(K)\frac{\mathop{\rm V}%
\nolimits(f^{\prime \prime })}{R({f^{\prime \prime }})R(f)}+{\ {0.72}} \right \}.
\end{eqnarray*}%
As in the last section, the second term is the only one which differs among bandwidth selectors. This second term was also calculated for the
quartic kernel, which is the kernel $K_{2r}$ with $r=2$. The calculation as above gave the value 1.89 and 0.83 instead of 2.19 and 0.72, see
Mammen et al.\ (2011).

The immediate lesson learned from comparing the asymptotic theory of
do-validation of the two kernels considered above is the following: the
second term is bigger for plug-in estimator for the quartic kernel than for the Epanechnikov estimator. However, for crossvalidation and do-validation it is the exact opposite, the second term is smaller for
the quartic kernel than for the Epanechnikov estimator.
Therefore, relatively speaking the validation approaches do better for the
higher order kernel $K_{2r}$ with $r=2$, than for the lower order kernel $K_{2r}$, with $r=1$ (the Epanechnikov kernel). One could argue that validation does better for the higher order kernel than for
the lower order kernel. However, lets further consider the case that we are really interested in
the optimal bandwidth for the lower order kernel and we really want
to use a validation approach to select that bandwidth, see Mammen et al.\ (2011)
for practical arguments for using validation
instead of plug-in. Then it seems intuitively appealing to carry that
validation out at the kernel with a high order to select the validated
bandwidth for that higher order kernel and then adjusting
this bandwidth to the lower order kernel by multiplying by
the kernel constant
\begin{equation} \label{Cind}
\left ( \frac{R(K)\mu_{2}^{2}(K_{2r})}{\mu_2^2(K) R(K_{2r})} \right )^{1/5}.
\end{equation}
And this is what we call indirect do-validation.

We now give a formal definition of the indirect do-validation bandwidth $ \widehat h_{IDO_r}$ with kernels $K$ and $K_{2r}$ as
\begin{equation} \label{r1}
	 \widehat h_{IDO_r} = \mathrm{C}_{I,r}  \widehat h_{DO,r}
\end{equation}
with $\widehat h_{DO,r}$ the do-validation bandwidth calculated with kernel $K_{2r}$, and $\mathrm{C}_{I,r}= \left ( \frac{R(K)\mu_{2}^{2}(K_{2r})}{\mu_2^2(K) R(K_{2r})} \right )^{1/5}$.
The (direct) do-validation bandwidth with kernel $K_{2r}$ is given by
\begin{equation} \label{r2}
\widehat h_{DO,r}=\mathrm{C}_{{r}}(\widehat{h}_{L,{r}}+\widehat{h}_{R,{r}}),
\end{equation}
where
\begin{equation} \label{r3}
\mathrm{C}_{{r}}=\left( \displaystyle\frac{R(K_{2r})}{\mu _{2}^{2}(K_{2r})}%
\displaystyle\frac{\mu _{2}^{2}(K_{L,{2r}})}{R(K_{L,{2r}})}\right) ^{1/5}. 
\end{equation}

Here $K_{L,2r}$ is defined from equation (\ref{L2}) replacing $K$ by $K_{2r}$. Also $\widehat{h}_{L,{2r}}$ and $\widehat{h}_{R,{2r}}$ are the crossvalidated bandwidths with kernels $K_{L,2r}$ and $K_{R,2r}$, respectively.
Now substituting (\ref{r2}) and (\ref{r3}) in (\ref{r1}) we get
\begin{equation} \label{r4}
\widehat	h_{IDO_r} = \left( \displaystyle\frac{R(K)}{\mu _{2}^{2}(K)}%
\displaystyle\frac{\mu _{2}^{2}(K_{L,{2r}})}{R(K_{L,{2r}})}\right) ^{1/5} (\widehat{h}_{L,{2r}}+\widehat{h}_{R,{2r}}).
\end{equation}

\begin{figure}[htb]
\begin{center}
 \includegraphics[width=11cm]{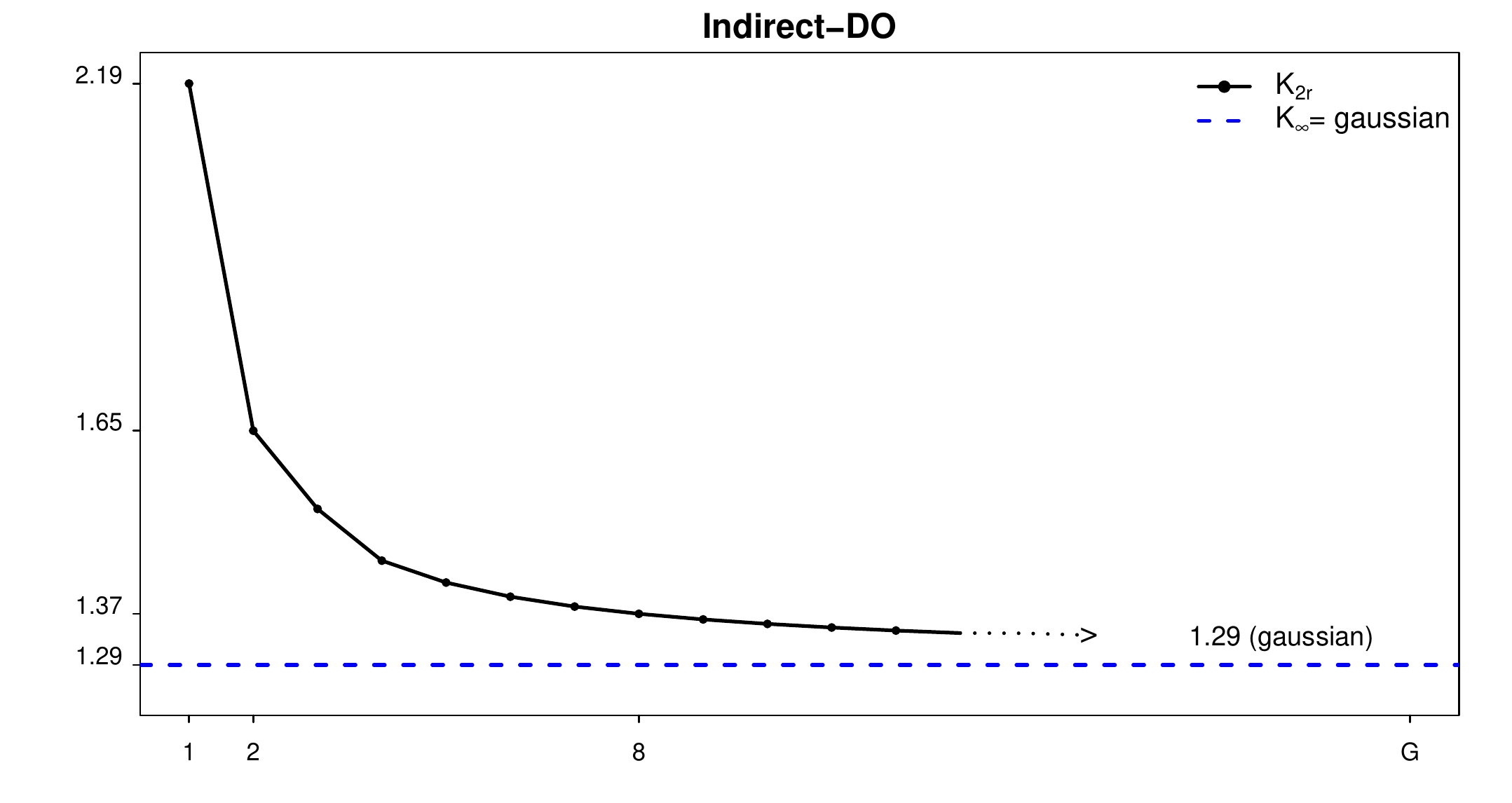}
\end{center}
 \vspace{-1cm}
\caption{\textit{Asymptotic variance term for indirect do-validation with kernels $K_{2r}$ when $r \to \infty$. The limit kernel is the Gaussian plotted with the discontinuous blue line.}}
\label{Fig:K2r}
\end{figure}

Using again Theorem 1 in Mammen et al.\ (2011) one gets that \begin{eqnarray}
n^{3/10}(\widehat{h}_{IDO_r}-h_{ISE})\rightarrow N(0,\sigma _{1,_{IDO}}^{2}) &&%
\mbox{in
distribution},  \label{asympeqB1}
\end{eqnarray}%
where $\sigma _{{IDO_r}}^{2}$
is given by
\begin{equation} \label{sigma:do2}
\begin{array}{rl}
\sigma_{IDO_r}^{2}= & \displaystyle\frac{4}{25}R(K)^{-2/5}\mu _{2}^{-6/5}(K)R({%
f^{\prime \prime }})^{-8/5}\mathop{\rm V}\nolimits(f^{\prime \prime })+\displaystyle\frac{1}{50}R(K)^{-7/5}\mu _{2}^{-6/5}(K) \\
& \ \ \times R({f^{\prime \prime }})^{-3/5}R(f)\displaystyle\int \left[
H(u)- \displaystyle\frac{%
R(K)}{R(K_{L,2r})} H_{IDO,r}(u)\right] ^{2}du
\end{array}
\end{equation}%
with $\mathop{\rm V}\nolimits(f^{\prime \prime })$ and $
H(u)$ as above and
\begin{eqnarray*}
H_{IDO,r}(d_{L,2r}u) &=&2\int K_{L,2r}(u+v)K_{L,2r}(v)dv+2\int K_{L,2r}(-u+v)K_{L,2r}(v)dv \\
&&+2\int K_{L,2r}(u+v)v K_{L,2r}^{\prime }(v)dv+2\int K_{L,2r}(-u+v)v K_{L,2r}^{\prime
}(v)dv \\
&&-2\left[ K_{L,2r}(u)+uK_{L,2r}^{\prime }(u)+K_{L,2r}(-u)-u K_{L,2r}^{\prime }(-u)\right]
, \\
&& \\
d_{L,2r} &=&\left( \frac{R(K)}{R(K_{L,2r})}\frac{\mu _{2}^{2}(K_{L,2r})}{\mu
_{2}^{2}(K)}\right) ^{-1/5}.\end{eqnarray*}
We get a result that is similar to the findings in our discussion of indirect crossvalidation in Section \ref{sec:icv}.
By increasing the order $r$ ($r=2,3,4,\ldots$) of the indirect kernel we get an incremental reduction in the asymptotic variance factor. Again, for $r\to \infty$ the factor converges to the factor of indirect do-validation with Gaussian kernel. This can be shown as in Section \ref{sec:icv}. Figure \ref{Fig:K2r} shows the factor as a function of $r$.

One sees that the trick of indirect
do-validation significantly improves on do-validation. Below we provide the resulting asymptotics for the indirect do-validation bandwidths, $h_{IDO_r}$,  with $r=1, 2,8$ and the Gaussian kernel, which is the limiting kernel as $r\to\infty$.

\begin{eqnarray*}
\sigma _{\mathop{\rm DO}\nolimits}^{2} &=&C_{f,K}\left\{ 4R(K)\frac{%
\mathop{\rm V}\nolimits(f^{\prime \prime })}{R({f^{\prime \prime }})R(f)}+{%
2.19}\right\} \\
\sigma _{\mathop{\rm IDO_2}\nolimits}^{2} &=&C_{f,K_{2}}\left\{ 4R(K)\frac{%
\mathop{\rm V}\nolimits(f^{\prime \prime })}{R({f^{\prime \prime }})R(f)}+{%
1.65}\right\} \\
\sigma _{\mathop{\rm IDO_8}\nolimits}^{2} &=&C_{f,K_{2}}\left\{ 4R(K)\frac{%
\mathop{\rm V}\nolimits(f^{\prime \prime })}{R({f^{\prime \prime }})R(f)}+{%
1.37}\right\} \\
\sigma _{\mathop{\rm IDO_{G}}\nolimits}^{2} &=&C_{f,K_{2}}\left\{ 4R(K)\frac{%
\mathop{\rm V}\nolimits(f^{\prime \prime })}{R({f^{\prime \prime }})R(f)}+{%
1.29}\right\} \\
\end{eqnarray*}%

\subsection{Simulation experiments about indirect do-validation}

\renewcommand{\baselinestretch}{1.2}
\begin{table}[hbt]
\begin{center}
{\small {\tabcolsep4.5pt
\begin{tabular}{c|rrrrr|rrrrr}
\multicolumn{1}{c}{} & \multicolumn{5}{c}{Design 1} & \multicolumn{5}{|c}{
Design 2} \\ \hline
\multicolumn{1}{c}{}  & {$h_{ISE}$}
& {\scriptsize {$\widehat h_{DO}$}} & {\scriptsize {$\widehat h_{IDO_2}$}} &
{\scriptsize {$\widehat h_{IDO_8}$}} & {\scriptsize {$\widehat h_{IDO_G}$}}
& {$h_{ISE}$}
& {\scriptsize {$\widehat h_{DO}$}} & {\scriptsize {$\widehat h_{IDO_2}$}} &
{\scriptsize {$\widehat h_{IDO_8}$}} & {\scriptsize {$\widehat h_{IDO_G}$}} \\
\hline
\multicolumn{1}{c}{}  & \multicolumn{10}{c}{$n=100$} \\ \hline
$m_1$ & 2.328  & 3.052 & 3.038 & 2.999 & 2.989 & 3.477 & 4.949 & 5.141 & 5.504 & 5.723\\
$m_2$ & 1.876  & 2.204 & 2.211 & 2.172 & 2.143 & 1.989 & 2.642 & 2.703 & 2.761 & 2.757\\
$m_3$ & 0.000  & 1.324 & 1.058 & 1.058 & 1.060 & 0.000 & 1.277 &1.385 &1.646 &1.832 \\
$m_4$ & 0.000  & 1.902 & 2.290 & 2.745 & 3.041 & 0.000 & 3.389 & 4.080 & 5.171 & 5.808 \\
$m_5$ & 0.000 & 0.583 & 0.603 & 0.616 & 0.633 & 0.000 & 0.914 & 0.999 &1.139 &1.204 \\ \hline
\multicolumn{1}{c}{}  & \multicolumn{10}{c}{$n=200$} \\ \hline
$m_1$ & 1.417  & 1.803 & 1.788 & 1.776 &1.775  & 2.307 & 2.930 & 2.925 & 3.011 & 3.108\\
$m_2$ & 1.098  & 1.402 & 1.373 & 1.341 &1.313  & 1.372 & 1.663 & 1.651 & 1.668 & 1.693\\
$m_3$ & 0.000  & 0.900 &0.833 &0.851 &0.880 & 0.000 & 0.748 & 0.755 & 0.893 & 1.029 \\
$m_4$ & 0.000  & 1.116 & 1.414 & 1.760 &2.022  & 0.000 & 1.607 & 1.865 & 2.421 & 2.859 \\
$m_5$ & 0.000 & 0.516 & 0.532 & 0.563 & 0.581 & 0.000 &  0.632 &0.667 &0.750 &0.826\\ \hline
\multicolumn{1}{c}{}  & \multicolumn{10}{c}{$n=500$} \\ \hline
$m_1$ & 0.731  & 0.903 & 0.889 & 0.878 & 0.876 & 1.208 & 1.439 & 1.439 & 1.442 & 1.458\\
$m_2$ & 0.465  & 0.559 & 0.553 & 0.537 & 0.532 & 0.648 & 0.775 & 0.773 & 0.771 & 0.777\\
$m_3$ & 0.000  & 0.750 & 0.690 & 0.667 & 0.688 & 0.000 & 0.526 &0.543 &0.568 &0.601 \\
$m_4$ & 0.000  & 0.418 & 0.618 & 0.836 & 0.990 & 0.000 & 0.679 & 0.832 & 1.058 & 1.258 \\
$m_5$ & 0.000 & 0.464 & 0.470 & 0.483 & 0.500 & 0.000 &   0.500& 0.524 &0.552 &0.601\\ \hline
\multicolumn{1}{c}{}  & \multicolumn{10}{c}{$n=1000$} \\ \hline
$m_1$ &0.439  & 0.525 & 0.519 & 0.514 & 0.513 & 0.732 & 0.846 & 0.841 & 0.839 & 0.842\\
$m_2$ &0.277  & 0.320 & 0.316 & 0.313 & 0.312 & 0.377 & 0.426 & 0.425 & 0.425 & 0.429\\
$m_3$ & 0.000 &  0.615& 0.535 &0.491 &0.480 & 0.000 &  0.459& 0.410& 0.377& 0.410 \\
$m_4$ &0.000  & 0.297 & 0.438 & 0.569 & 0.659 & 0.000 & 0.345 & 0.423 & 0.564 & 0.681 \\
$m_5$ & 0.000 & 0.434 & 0.432 & 0.449 & 0.464 & 0.000 & 0.421 & 0.428 & 0.448 & 0.471 \\ \hline
\end{tabular}
}}
\end{center}
\caption{\textit{Simulation results about the indirect do-validation method with designs 1 and 2. We compare the original do-validated bandwidth, $\widehat h_{DO}$, with three indirect versions $\widehat h_{IDO_2}$, $\widehat h_{IDO_8}$ and $\widehat h_{IDO_G}$ for kernels $K_{2r}$ with $r=2,8, \infty$.}}
\label{tab-IDO-1}
\end{table}

\renewcommand{\baselinestretch}{1.2}
\begin{table}[hbt]
\begin{center}
{\small {\tabcolsep4.5pt
\begin{tabular}{c|rrrrr|rrrrr}
\multicolumn{1}{c}{} & \multicolumn{5}{c}{Design 3} & \multicolumn{5}{|c}{
Design 4} \\ \hline
\multicolumn{1}{c}{}  & {$h_{ISE}$}
& {\scriptsize {$\widehat h_{DO}$}} & {\scriptsize {$\widehat h_{IDO_2}$}} &
{\scriptsize {$\widehat h_{IDO_8}$}} & {\scriptsize {$\widehat h_{IDO_G}$}}
& {$h_{ISE}$}
& {\scriptsize {$\widehat h_{DO}$}} & {\scriptsize {$\widehat h_{IDO_2}$}} &
{\scriptsize {$\widehat h_{IDO_8}$}} & {\scriptsize {$\widehat h_{IDO_G}$}} \\
\hline
\multicolumn{1}{c}{}  & \multicolumn{10}{c}{$n=100$} \\ \hline
$m_1$ &4.448  & 11.283 & 11.597 & 11.532 & 11.328 & 4.842 & 6.462 & 6.483 & 6.536 & 6.601\\
$m_2$ &2.231  & 3.885 & 3.774 &  3.904 &  3.960 & 2.644& 3.246 & 3.209 & 3.208 & 3.216\\
$m_3$ &0.000  & 4.913 & 4.996 & 4.913 & 4.782 & 0.000 &  0.943 &0.941 &1.021 &1.027 \\
$m_4$ &0.000  & 10.198 & 10.687 & 10.793 & 10.661 & 0.000 & 3.352 & 3.619 & 4.019 & 4.300 \\
$m_5$ & 0.000 &  1.943 & 1.943 & 1.943 & 1.943 & 0.000 & 0.894 &0.944 &1.000& 1.027\\  \hline
\multicolumn{1}{c}{}  & \multicolumn{10}{c}{$n=200$} \\ \hline
$m_1$ &2.830  & 3.799 & 3.997 & 4.207 & 4.391 & 3.100 & 3.940 & 3.955 & 3.956&3.984\\
$m_2$ &1.343  & 2.288 & 2.560 & 2.624 & 2.630 & 1.657 & 2.032 & 2.018 & 1.983&1.980\\
$m_3$ &0.000  & 0.811 & 0.941 & 1.256 &1.368 & 0.000 &  0.774 &0.797 &0.830 &0.871 \\
$m_4$ &0.000  & 1.895 & 2.307 & 2.899 & 3.345 & 0.000 & 2.147 & 2.371 & 2.670&2.904 \\
$m_5$ & 0.000 & 0.734 & 0.896 & 1.085 & 1.159 & 0.000 & 0.794 & 0.853 &0.912 &0.922\\ \hline
\multicolumn{1}{c}{}  & \multicolumn{10}{c}{$n=500$} \\ \hline
$m_1$ &1.540  & 1.757 & 1.751 & 1.767 & 1.798 & 1.687 & 1.967 & 1.956 & 1.961 & 1.974\\
$m_2$ &0.685  & 0.815 & 0.806 & 0.808 & 0.829 & 0.767 & 0.882 & 0.877 & 0.873 & 0.870\\
$m_3$ &0.000  & 0.397 & 0.390 & 0.417 &0.458 & 0.000 &  0.491 &0.456 &0.511 &0.548 \\
$m_4$ &0.000  & 0.545 & 0.627 & 0.839 & 1.036 & 0.000 & 0.973 & 1.108 & 1.368 & 1.546 \\
$m_5$ & 0.000 & 0.438 & 0.440 & 0.498 & 0.531 & 0.000 & 0.587 & 0.587 & 0.617 & 0.632\\ \hline
\multicolumn{1}{c}{}  & \multicolumn{10}{c}{$n=1000$} \\ \hline
$m_1$ & 0.943 & 1.044 & 1.039 & 1.039 & 1.045 & 1.060 & 1.174 & 1.169 & 1.175 & 1.183\\
$m_2$ & 0.405 & 0.449 & 0.450 & 0.454 & 0.462 & 0.491 & 0.534 & 0.523 & 0.518 & 0.517\\
$m_3$ &0.000  & 0.300 & 0.274 & 0.270 & 0.281 & 0.000 &  0.322& 0.315 &0.326 &0.345 \\
$m_4$ & 0.000 & 0.206 & 0.279 & 0.408 & 0.517 & 0.000 & 0.544 & 0.662 & 0.870 & 1.008 \\
$m_5$ & 0.000 & 0.368 & 0.368 & 0.400 & 0.435 & 0.000 & 0.470 & 0.470 & 0.498 & 0.502\\ \hline
\end{tabular}
}}
\end{center}
\caption{\textit{Simulation results about the indirect do-validation method with designs 3 and 4. We compare the original do-validated bandwidth, $\widehat h_{DO}$, with three indirect versions $\widehat h_{IDO_2}$, $\widehat h_{IDO_8}$ and $\widehat h_{IDO_G}$ for kernels $K_{2r}$ with $r=2,8, \infty$.}}
\label{tab-IDO-2}
\end{table}

\renewcommand{\baselinestretch}{1.2}
\begin{table}[hbt]
\begin{center}
{\small {\tabcolsep4.5pt
\begin{tabular}{c|rrrrr|rrrrr}
\multicolumn{1}{c}{} & \multicolumn{5}{c}{Design 5} & \multicolumn{5}{|c}{
Design 6} \\ \hline
\multicolumn{1}{c}{}  & {$h_{ISE}$}
& {\scriptsize {$\widehat h_{DO}$}} & {\scriptsize {$\widehat h_{IDO_2}$}} &
{\scriptsize {$\widehat h_{IDO_8}$}} & {\scriptsize {$\widehat h_{IDO_G}$}}
& {$h_{ISE}$}
& {\scriptsize {$\widehat h_{DO}$}} & {\scriptsize {$\widehat h_{IDO_2}$}} &
{\scriptsize {$\widehat h_{IDO_8}$}} & {\scriptsize {$\widehat h_{IDO_G}$}} \\
\hline
\multicolumn{1}{c}{}  & \multicolumn{10}{c}{$n=100$} \\ \hline
$m_1$ & 3.356 & 4.437 & 4.509 & 4.533 & 4.539 & 3.633 & 4.972 & 5.036 & 5.136 & 5.211\\
$m_2$ & 1.383 & 1.566 & 1.534 & 1.476 & 1.446 & 1.617 & 1.911 & 1.884 & 1.850 & 1.817\\
$m_3$ & 0.000 & 1.061 &1.137& 1.185& 1.185 & 0.000 &  1.000& 1.031& 1.078& 1.157 \\
$m_4$ & 0.000 & 5.434 & 6.041 & 6.557 & 6.721 & 0.000 & 5.794 & 6.305 & 6.913 & 7.317 \\
$m_5$ & 0.000 & 0.999 & 0.999 & 1.021 & 1.042 & 0.000 & 1.042 & 1.084 & 1.131 &1.174 \\ \hline
\multicolumn{1}{c}{}  & \multicolumn{10}{c}{$n=200$} \\ \hline
$m_1$ &2.293  & 3.008 & 3.036 & 3.054 & 3.063 & 2.387 & 3.206 & 3.250 & 3.310 & 3.362\\
$m_2$ &0.864  & 1.002 & 0.984 & 0.918 & 0.896 & 0.955 & 1.242 & 1.270 & 1.272 & 1.273\\
$m_3$ & 0.000 & 1.029& 1.050& 1.048& 1.065 & 0.000 & 0.971& 1.006& 1.052& 1.086 \\
$m_4$ &0.000  & 4.331 & 4.707 & 5.149 & 5.378 & 0.000 & 4.163 & 4.508 & 5.018 & 5.386 \\
$m_5$ &0.000 & 0.923 & 0.923 & 0.977 & 0.999 & 0.000 & 0.975 & 0.999 & 1.067 &1.090 \\ \hline
\multicolumn{1}{c}{}  & \multicolumn{10}{c}{$n=500$} \\ \hline
$m_1$ &1.287  & 1.668 & 1.678 & 1.695 & 1.710 & 1.355 & 1.694 & 1.700 & 1.718 & 1.738\\
$m_2$ &0.520  & 0.550 & 0.537 & 0.520 & 0.512 & 0.495 & 0.621 & 0.617 & 0.604 & 0.596\\
$m_3$ & 0.000 &  0.911& 0.958& 0.990& 1.027 & 0.000 &  0.712& 0.714& 0.709& 0.750 \\
$m_4$ &0.000  & 3.211 & 3.449 & 3.820 & 4.046 & 0.000 & 2.430 & 2.624 & 2.935 & 3.148 \\
$m_5$ & 0.000 &  0.875& 0.928 &0.951  & 0.976 & 0.000 &  0.751& 0.751 & 0.786 &0.800\\ \hline
\multicolumn{1}{c}{}  & \multicolumn{10}{c}{$n=1000$} \\ \hline
$m_1$ & 0.844 & 1.016 & 1.023 & 1.036 & 1.050 & 0.892 & 1.030 & 1.031 & 1.043 & 1.056\\
$m_2$ & 0.357 & 0.396 & 0.393 & 0.381 & 0.374 & 0.304 & 0.334 & 0.327 & 0.318 & 0.315\\
$m_3$ & 0.000 &  0.612& 0.632& 0.717& 0.771 & 0.000 &  0.509& 0.480& 0.492& 0.543 \\
$m_4$ & 0.000 & 1.868 & 2.064 & 2.387 & 2.605 & 0.000 & 1.448 & 1.580 & 1.878 & 2.077 \\
$m_5$ & 0.000 & 0.718 & 0.728 & 0.775 & 0.799 & 0.000 & 0.600 & 0.595 & 0.638 & 0.684\\ \hline
\end{tabular}
}}
\end{center}
\caption{\textit{Simulation results about the indirect do-validation method with designs 5 and 6. We compare the original do-validated bandwidth, $\widehat h_{DO}$, with three indirect versions $\widehat h_{IDO_2}$, $\widehat h_{IDO_8}$ and $\widehat h_{IDO_G}$ for kernels $K_{2r}$ with $r=2,8, \infty$.}}
\label{tab-IDO-3}
\end{table}

Here we extend the simulation experiments carried out for indirect crossvalidation above with the just defined  indirect do-validation method. We evaluate the finite sample performance of highering the orders of the indirect kernel for do-validation and compare with the former do-validation and the optimal ISE bandwidth ($h_{ISE}$). We consider in the study three possible indirect do-validation bandwidths: $\widehat h_{IDO_2}$, $\widehat h_{IDO_8}$ and $\widehat h_{IDO_G}$, which comes from using as the kernel $K$ the Epanechnikov kernel and as kernel $L$ the higher order polynomial kernel, $K_{2r}$ defined in (\ref{def:K2r}), for $r=2,8, \infty$ (with $K_{\infty}$ being the Gaussian kernel). Again we consider the six density estimation problems showed in Figure \ref{fig-dgp} and the five performance measures defined in (\ref{m1})-(\ref{m5}).

Tables \ref{tab-IDO-1}, \ref{tab-IDO-2} and \ref{tab-IDO-3} show the simulation results. As we saw for indirect classical  crossvalidation, the finite sample bias ($m_4$) is consistently increasing when highering the order of the indirect kernel. However, this increase in bias is offset by a decrease in volatility ($m_2$). This is consistently over sample size and design and follow the results we saw in the previous section for classical crossvalidation. However, when it comes to the overall average integrated squared error performance the impression is less clear. Sometimes increasing the order of the indirect kernel improves results, sometimes it does not. Overall, the indirect do-validation methods perform more or less the same. Therefore, for do-validation the decrease in volatility (m2) and the increase ($m_4$) seem to be effects of similar size overall. So, the estimators have similar averaged ISE behavior, but they are quite different, when it comes to their bias/variance trade offs. Therefore, also for indirect do-validation it looks quite promising to take in a collection of indirect do-validated bandwidth selectors and use them for our median bandwidth selector that will be introduced in the next section.

\section{A comparison to plug-in density estimation}
\label{sec:pi}

In this section, for completeness, we first state asymptotics for the plug-in bandwidth selector. It holds for symmetric kernel $K$ that \begin{eqnarray}
n^{3/10}(\widehat{h}_{PI}-h_{ISE})\rightarrow N(0,\sigma_{PI}^{2}) &&%
\mbox{in
distribution},  \label{asympeq1}
\end{eqnarray}%
where
\begin{equation} \label{sigma:pi}
\begin{array}{rl}
\sigma_{PI}^{2}= & \displaystyle\frac{4}{25}R(K)^{-2/5}\mu _{2}^{-6/5}(K)R({%
f^{\prime \prime }})^{-8/5}\mathop{\rm V}\nolimits(f^{\prime \prime })+\displaystyle\frac{1}{50}R(K)^{-7/5}\mu _{2}^{-6/5}(K) \\
& \ \ \times R({f^{\prime \prime }})^{-3/5}R(f)\displaystyle\int
H(u)^{2}du,%
\end{array}
\end{equation}%
with $\mathop{\rm V}\nolimits(f^{\prime \prime })$ and $H(u)$ as above,
see e.g.\ Mammen et al.\ (2011).

The simulation results for the plug-in method will be described in the next subsection at the same time as we introduce a new bandwidth estimator.  To allow for direct comparison with Mammen et al.\ (2011) we have implemented their refined plug-in estimator which followed the proposals of Sheather and Jones (1991) and Park and Marron (1990). We refer the reader to this former paper for more details.

\renewcommand{\baselinestretch}{1.2}
\begin{table}[hbt]
\begin{center}
{\small {\tabcolsep4.5pt
\begin{tabular}{c|rrr|rrr|rrr}
\multicolumn{1}{c}{} & \multicolumn{3}{c}{Design 1} & \multicolumn{3}{|c}{
Design 2}  & \multicolumn{3}{|c}{Design 3} \\ \hline
\multicolumn{1}{c}{}  & {$h_{ISE}$} & {\scriptsize {$\widehat h_{PI}$}} &  median
& {$h_{ISE}$} & {\scriptsize {$\widehat h_{PI}$}} & median
& {$h_{ISE}$} & {\scriptsize {$\widehat h_{PI}$}} & median \\ \hline
\multicolumn{1}{c}{}  & \multicolumn{9}{c}{$n=100$} \\ \hline
$m_1$ & 2.328 & 2.905 &2.996& 3.477 & 7.703 &5.488& 4.448 & 14.022 &11.514 \\
$m_2$ & 1.876 & 2.034 &2.162& 1.989 & 1.737 &2.733& 2.231 &  0.797 & 3.905 \\
$m_3$ & 0.000 & 0.972 &1.066& 0.000 & 4.087 &1.626& 0.000 &  5.995 & 4.913  \\
$m_4$ & 0.000 & 3.467 &2.864& 0.000 & 9.654 &5.189& 0.000 & 14.317 &11.8005  \\
$m_5$ & 0.000 & 0.666 &0.666& 0.000 & 1.301 &1.136& 0.000 & 1.943 & 1.943 \\ \hline
\multicolumn{1}{c}{}  & \multicolumn{9}{c}{$n=200$} \\ \hline
$m_1$ & 1.417 & 1.725 &1.782& 2.307 & 4.916 &3.046& 2.830 & 12.034 &4.204 \\
$m_2$ & 1.098 & 1.215 &1.338& 1.372 & 1.360 &1.678& 1.343 &  0.657 &2.603 \\
$m_3$ & 0.000 & 0.750 &0.839& 0.000 & 3.410 &0.958& 0.000 &  7.391 &1.239  \\
$m_4$ & 0.000 & 2.814 &2.026& 0.000 & 7.498 &2.582& 0.000 & 12.857 &2.934  \\
$m_5$ & 0.000 & 0.581 &0.564& 0.000 & 1.212 &0.790& 0.000 &  2.126 &1.071 \\ \hline
\multicolumn{1}{c}{}  & \multicolumn{9}{c}{$n=500$} \\ \hline
$m_1$ & 0.731 & 0.857 &0.884& 1.208 & 2.408 &1.470& 1.540 & 7.752 &1.791 \\
$m_2$ & 0.465 & 0.513 &0.540& 0.648 & 0.858 &0.784& 0.685 & 0.729 &0.819 \\
$m_3$ & 0.000 & 0.569 &0.667& 0.000 & 2.295 &0.617& 0.000 & 9.107 &0.458  \\
$m_4$ & 0.000 & 1.877 &1.125& 0.000 & 5.330 &1.277& 0.000 & 9.629 &0.967  \\
$m_5$ & 0.000 & 0.518 &0.518& 0.000 & 1.052 &0.632& 0.000 & 1.998 &0.532 \\ \hline
\multicolumn{1}{c}{}  & \multicolumn{9}{c}{$n=1000$} \\ \hline
$m_1$ & 0.439 & 0.506 &0.519& 0.732 & 1.331 &0.857& 0.943 & 4.918 &1.052 \\
$m_2$ & 0.277 & 0.311 &0.316& 0.377 & 0.559 &0.440& 0.405 & 0.628 &0.464 \\
$m_3$ & 0.000 & 0.474 &0.533& 0.000 & 1.787 &0.479& 0.000 & 8.447 &0.319  \\
$m_4$ & 0.000 & 1.437 &0.850& 0.000 & 4.001 &0.783& 0.000 & 7.699 &0.539  \\
$m_5$ & 0.000 & 0.483 &0.483& 0.000 & 0.946 &0.473& 0.000 & 1.801 &0.435 \\ \hline
\end{tabular}
}}
\end{center}
\caption{\textit{Simulation study about the plug-in method and the median estimator for designs 1 to 3.}}
\label{tab-IPI-1}
\end{table}

\subsection{Combination of bandwidth selectors: a median estimator}

\renewcommand{\baselinestretch}{1.2}
\begin{table}[hbt]
\begin{center}
{\small {\tabcolsep4.5pt
\begin{tabular}{c|rrr|rrr|rrr}
\multicolumn{1}{c}{} & \multicolumn{3}{c}{Design 4} & \multicolumn{3}{|c}{
Design 5}  & \multicolumn{3}{|c}{Design 6} \\ \hline
\multicolumn{1}{c}{}  & {$h_{ISE}$} & {\scriptsize {$\widehat h_{PI}$}} &  median
& {$h_{ISE}$} & {\scriptsize {$\widehat h_{PI}$}} & median
& {$h_{ISE}$} & {\scriptsize {$\widehat h_{PI}$}} & median \\ \hline
\multicolumn{1}{c}{}  & \multicolumn{9}{c}{$n=100$} \\ \hline
$m_1$ & 4.842 & 8.259 &6.544& 3.356 & 4.623 &4.536& 3.633 & 5.648 &5.143 \\
$m_2$ & 2.644 & 3.544 &3.216& 1.383 & 1.362 &1.468& 1.617 & 1.471 &1.847 \\
$m_3$ & 0.000 & 1.962 &1.010& 0.000 &1.220  &1.215& 0.000 & 1.415 &1.084 \\
$m_4$ & 0.000 & 7.543 &4.095& 0.000 & 7.681 &6.581& 0.000 & 9.586 &6.973 \\
$m_5$ & 0.000 & 1.316 &1.001& 0.000 & 1.084 &1.042& 0.000 & 1.174 &1.136 \\ \hline
\multicolumn{1}{c}{}  & \multicolumn{9}{c}{$n=200$} \\ \hline
$m_1$ & 3.100 & 5.250 &3.964& 2.293 & 3.394 &3.056& 2.387 & 4.299 &3.318 \\
$m_2$ & 1.657 & 2.176 &1.986& 0.864 & 0.805 &0.917& 0.955 & 0.988 &1.271 \\
$m_3$ & 0.000 & 1.818 &0.831& 0.000 & 1.348 &1.048& 0.000 & 1.772 &1.059  \\
$m_4$ & 0.000 & 6.298 &2.750& 0.000 & 7.629 &5.191& 0.000 & 9.620 &5.071  \\
$m_5$ & 0.000 & 1.223 &0.917& 0.000 & 1.130 &0.977& 0.000 & 1.380 &1.070\\ \hline
\multicolumn{1}{c}{}  & \multicolumn{9}{c}{$n=500$} \\ \hline
$m_1$ & 1.687 & 2.797 &1.970& 1.287 & 2.200 &1.697& 1.355 & 2.956 &1.721 \\
$m_2$ & 0.767 & 1.058 &0.872& 0.520 & 0.474 &0.521& 0.495 & 0.510 &0.603 \\
$m_3$ & 0.000 & 1.480 &0.533& 0.000 & 1.689 &1.000& 0.000 & 2.348 &0.712  \\
$m_4$ & 0.000 & 4.828 &1.455& 0.000 & 7.253 &3.848& 0.000 & 9.022 &2.970  \\
$m_5$ & 0.000 & 1.058 &0.617& 0.000 & 1.229 &0.952& 0.000 & 1.380 &0.786\\ \hline
\multicolumn{1}{c}{}  & \multicolumn{9}{c}{$n=1000$} \\ \hline
$m_1$ & 1.060 & 1.757 &1.183& 0.844 & 1.531 &1.038& 0.892 & 2.207 &1.045 \\
$m_2$ & 0.491 & 0.631 &0.518& 0.357 & 0.318 &0.380& 0.304 & 0.344 &0.317 \\
$m_3$ & 0.000 & 1.462 &0.352& 0.000 & 1.886 &0.721& 0.000 & 2.891 &0.504  \\
$m_4$ & 0.000 & 4.099 &0.970& 0.000 & 6.356 &2.427& 0.000 & 8.322 &1.907  \\
$m_5$ & 0.000 & 1.002 &0.528& 0.000 & 1.191 &0.775& 0.000 & 1.630 &0.650\\ \hline
\end{tabular}
}}
\end{center}
\caption{\textit{Simulation study about the plug-in method and the median estimator for designs 4 to 6.}}
\label{tab-IPI-2}
\end{table}

In this section we take advantage of the lessons learned in the above sections on indirect crossvalidation, indirect do-validation and indirect plug-in estimation. We learned that all these estimators were different giving us the idea  that a median of estimators might perform very well. In this section we define the median estimator as the median of  13 bandwidth values, where 8 are crossvalidated bandwidths and 5 are identical values that are equal to the outcome of the plug-in estimator. The crossvalidated bandwidths are
  our four choices of cross-validated bandwidths and our four choices of do-validated bandwidths.  We also tried other combinations. The median here had the best performance with other combinations being very close in performance.   A comparison between the plug-in method and our median estimator is provided in Tables \ref{tab-IPI-1} and \ref{tab-IPI-2}. As a benchmark we include the ISE optimal bandwidth. The median behaves very well on all measures. However, the performance  of the median estimator is so close to the performance of the do-validated estimator that we finally prefer the latter do-validated estimator in the end. The do-validated estimator is simpler to calculate and simpler to generalize to more complicated settings.

\section*{References}

\begin{description}

%
%
%

\item Cheng, M.Y., 1997a. Boundary-aware estimators of integrated squared
density derivatives. \emph{Journal of the Royal Statistical Society Ser. B},
{\bfseries 50}, 191--203.

\item Cheng, M.Y., 1997b. A bandwidth selector for local linear density
estimators. \emph{The Annals of Statistics}, {\bfseries 25}, 1001--1013.
%

\item  G\'amiz-P\'erez, M.L., Mart{\'\i}nez-Miranda, M.D., Nielsen, J.P., 2012. Smoothing survival densities in practice. To appear in \emph{Computational Statistics and Data Analysis}.


%

%
%


\item Hart, J.D., Lee, C.-L., 2005. Robustness of one-sided
cross-validation to autocorrelation. \emph{Journal of Multivariate Statistics%
}, {\bfseries 92}, 77--96.

\item Hart, J.D., Yi, S., 1998. One-Sided Cross-Validation. \emph{Journal
of the American Statistical Association}, {\bfseries 93}, 620--631.

\item Jones, M.C., 1993. Simple boundary correction in kernel density
estimation. \emph{Statistics and Computing}, {\bfseries 3}, 135--146.


\item Mammen, E., Mart{\'\i}nez-Miranda, M.D., Nielsen, J.P., Sperlich, S., 2011. {Do-validation for kernel density estimation}, \emph{Journal of the American Statistical Association}, {\bfseries 106}, 651--660.

\item Mart{\'\i}nez-Miranda, M.D., Nielsen, J.P., Sperlich, S., 2009. One
sided crossvalidation for density estimation with an application to
operational risk. \emph{In "Operational Risk Towards Basel III: Best
Practices and Issues in Modelling. Management and Regulation, ed. G.N.\
Gregoriou; John Wiley and Sons, Hoboken, New Jersey.}

\item  Oliveira, M., Crujeiras, R.M., Rodríguez-Casal, A., 2012. A plug-in rule for bandwidth selection in circular density estimation, \emph{ Computational Statistics and Data Analysis}, {\bfseries56 }(12), 3898--3908.

\item Park, B.U., Marron, J.S., 1990. Comparison of Data-Driven Bandwidth
Selectors. \emph{Journal of the American Statistical Association}, {%
\bfseries 85}, 66--72.


\item Savchuk, O.Y., Hart, J.D., Sheather S.J., 2010a. Indirect
crossvalidation for Density Estimation. \emph{Journal of the American
Statistical Association}, {\bfseries 105}, 415--423.

\item Savchuk, O.Y., Hart, J.D., Sheather S.J., 2010b. An empirical
study of indirect crossvalidation for Density Estimation. \emph{IMS Lecture
Notes - Festschrift for Tom Hettmansperger.}


\item Sheather, S.J., Jones, M.C., 1991. A reliable data-based bandwidth
selection method for kernel density estimation. \emph{Journal of the Royal
Statistical Society}, Ser. B, {\bfseries53}, 683--690.

\item  Soni, P., Dewan, I., Jain, K., 2012.  Nonparametric estimation of quantile density function. \emph{ Computational Statistics and Data Analysis}, {\bfseries56 }(12), 3876–-3886.

%
%
\end{description}

\end{document}